\documentstyle[12pt]{article}
\newcommand{\beq}{\begin{equation}}
\newcommand{\eeq}{\end{equation}}
\newcommand{\beqa}{\begin{eqnarray}}
\newcommand{\eeqa}{\end{eqnarray}}
\newcommand{\beqan}{\begin{eqnarray*}}
\newcommand{\eeqan}{\end{eqnarray*}}
\newcommand{\no}{\nonumber}

\newcommand{\ul}{\underline}
\newcommand{\ol}{\overline}
\newcommand{\ra}{\rightarrow}

\newcommand{\Lra}{\Longrightarrow}

\newcommand{\ben}{\begin{enumerate}}
\newcommand{\een}{\end{enumerate}}
\newcommand{\bfl}{\begin{flushleft}}
\newcommand{\efl}{\end{flushleft}}
\newcommand{\ba}{\begin{array}}
\newcommand{\ea}{\end{array}}
\newcommand{\btab}{\begin{tabular}}
\newcommand{\etab}{\end{tabular}}
\newcommand{\bit}{\begin{itemize}}
\newcommand{\eit}{\end{itemize}}

\newcommand{\A}{{\cal A}}
\newcommand{\B}{{\cal B}}

\newcommand{\vs}{\vspace}
\newcommand{\hs}{\hspace}
\newcommand{\ld}{\lambda}
\newcommand{\prepr}[1] {\begin{flushright} {\bf #1} \end{flushright}
\vskip
1.5cm}
\newcommand{\titul}[1] {\begin{center} \Large{ }{\bf #1 }
\end{center}\vskip 1.cm}

\newcommand{\autor}[1] {\begin{center} {\bf \lineskip .3cm #1  }
                        \end{center} }

\newcommand{\lugar}[1] {\begin{center}  {\large \it #1   }
\end{center}}
\newcommand{\abstr}[1] {{\begin{center} \vskip .5cm {\large \bf
Abstract
                        \vspace{0pt}} \end{center}}\begin{quote} #1
                        \end{quote}}
\newcounter{muni}

\begin{document}
\vspace{.4cm}
\hbadness=10000
\pagenumbering{arabic}
\begin{titlepage}
\prepr{Preprint hep-ph/9409417 \\PAR/LPTHE/94-40  \\ 05 Octorber, 1994 \\
(Revised Version) }
\titul{ On the determination of the  relative sign \\
of  $a_1$ and $a_2$  from  polarization measurements \\
in ${\bf B_u^- \ra \rho^- {D^*}^o }$ decay.}
\autor{
 Y. Y. KEUM\footnote{\rm Postal address: LPTHE, Tour 24,
$5^{\grave{e}me}$ Etage, Universit\'e Pierre {\it \&} Marie Curie and
Universit\'e Denis Diderot, 2 Place Jussieu, F-75251 Paris CEDEX 05,
France.} }

\lugar{Laboratoire de Physique Th\'eorique et Hautes
Energies,\footnote{\em Unit\'e associ\'ee au CNRS URA 280}
 Paris, France }

\vs{-12cm}
\thispagestyle{empty}
\vs{120mm}
\noindent
\abstr{
We point out that polarization measurements such as the longitudinal fraction
and the transverse left-right asymmetry in the Bauer-Stech-Wirbel (BSW)
class III processes involving two final vector mesons,
$B^- \ra \rho^- {D^*}^o$ taken as an example,
are useful in determining the relative sign as well as the relative
magnitudes of the coefficients $a_1$ and $a_2$.

 }
\vs{40mm}
\begin{center}
{\bf  E-mail :  keum@lpthe.jussieu.fr }
\end{center}
\end{titlepage}


\newpage
\hs{3mm} \Large{ } {\bf I. \hs{3mm} Introduction}
\vs{6mm} \normalsize

During the last few years, there has been considerable interest
[1 - 7] in the determination of the phenomenological
free parameters $a_1$ and $a_2$ introduced by Bauer, Stech and Wirbel
(BSW henceforth) \cite{BSW} in $B$ meson decays.
{}From a phenomenological point of view, there is no constraint
on these parameters. We take them as  free parameters
to be determined from experiments.\\

However, there are some ambiguities in the sign of $a_2/a_1$ in $B$
meson decays .
Among two possible solutions, positive or negative sign
for $a_2/a_1$ as was found in Ref.\cite{Neubert},
 recent CLEO II data \cite{{Browder},{CLEO94-5},{Yamamoto}} indicate that
$a_2/a_1$ is positive.

\vs{2mm}
In this paper, we suggest a new method to determine
the size and the sign of $a_2/a_1$.
Since the decay amplitudes of the Class III depend on both $a_1$ and
$a_2$,
measurements of the longitudinal polarization fraction
and the left-right asymmetry  in transverse polarization
for $B^- \ra \rho^- {D^*}^o$, which has a sizeable branching ratio,
can provide a way to settle the question. \\

\vs{3mm} $1^{\bullet})$ \hs{2mm}
We start by recalling
the relevant effective weak Hamiltonian :
\beq
{\cal H}_{eff} = \frac{G_F}{\sqrt{2}} \ V_{cb} V_{ud}^* [ C_1(\mu)
{\cal O}_1 + C_2(\mu) {\cal O}_2 ] \hs{2mm} + \hs{2mm} H.C.
\label{Heff}
\eeq
\beq
{\cal O}_{1} = (\ol{d} \Gamma^{\rho} u)(\ol{c} \Gamma_{\rho}b),
\hs{20mm}
{\cal O}_{2} = (\ol{c} \Gamma^{\rho} u)(\ol{d} \Gamma_{\rho} b)
\label{Op}
\eeq
where $G_F $ is the Fermi coupling constant,
$V_{cb}$ and $V_{ud}$ are the relevant Cabibbo-Kobayashi-Maskawa
(CKM) matrix elements
and $\Gamma_{\rho} = \gamma_{\rho}(1 - \gamma_5)$.

The Wilson coefficients $C_1(\mu)$ and $C_2(\mu)$ incorporate the
short-distance effects arising from the renormalization of ${\cal
H}_{eff}$ from $\mu = m_W$ to $\mu = O(m_b)$. These coefficients are
known to next-to-leading order (NLL) \cite{{NLL1},{NLL2}}.

In the leading-logarithm approximation (LLA) \cite{LLA}, we have
\beq
C_1 = \frac{1}{2} \ ( C_{+} + C_{-} ), \hs{10mm} C_2 = \frac{1}{2} \
( C_{+} - C_{-} ) \label{Coff}
\eeq
with

\beq
C_{\pm} = [ \frac{\alpha_s(\mu^2)}{\alpha_s(M_W^2)} \ ]^{ \frac{2
\gamma_{\pm}}{b} \
 } \label{Cpm}
\eeq
where $\gamma_{-} = - 2 \gamma_{+} = 2 $, and $\alpha_s(\mu^2)$ is
the running coupling constant of the strong interaction given by

\beq
\alpha_s(\mu^2) = \frac{4 \pi}{ b \log( \mu^2/ \Lambda^2_{QCD} ) } \
\hs{15mm} {\rm with} \hs{5mm} b = 11 - \frac{2}{3} \ n_f \label{AS}
\eeq
where $n_f$ being the number of 'active' flavours and $\mu$ is the
typical mass scale of the problem.

With $\mu = m_b = 5 $ GeV, $n_f = 4 $ and $\Lambda_{QCD} = 0.25$ GeV,
we have

\beq
C_1({\rm LLA}) = 1.11, \hs{20mm} C_2({\rm LLA}) = - 0.26 \label{Cval}
\eeq

Including next-to-leading log corrections one obtains \cite{Ruckl}

\beq
C_1({\rm LLA + NNL}) = 1.13 \hs{20mm} C_2({\rm LLA + NLL}) = - 0.29
\label{CNNL}
\eeq
i.e. a very mild enhancement of the original charged current coupling
together with an induced neutral current operator with moderate
strength.

After Fierz transformation, ${\cal O}_1 $ and ${\cal O}_2 $ can be
written as
\cite{Fierz}
\beqa
{\cal O}_1 &=& \frac{1}{N_c} \ {\cal O}_2 + 2 \tilde{\cal O}_2
\hs{15mm}
\tilde{\cal O}_2 = (\ol{d} \Gamma^{\rho} \frac{\lambda^a}{2} \
b)(\ol{c} \Gamma_{\rho} \frac{\lambda^a}{2} \ u) \label{eqO1} \\
{\cal O}_2 &=& \frac{1}{N_c} \ {\cal O}_1 + 2 \tilde{\cal O}_1
\hs{15mm}
\tilde{\cal O}_1 = (\ol{d} \Gamma^{\rho} \frac{\lambda^a}{2} \
u)(\ol{c} \Gamma_{\rho} \frac{\lambda^a}{2} \ b) \label{eqO2}
\eeqa
where $N_c = 3$ is the number of colours
and the second terms, $\tilde{\cal O}_1$ and $\tilde{\cal O}_2$,
sometimes called
non-factorizable terms, are composed of  two color-octet currents
with $\lambda^a$ being
the colour SU(3) Gell-Mann matrices.

We can rewrite Eq.(1) in two ways
\beqa
C_1 {\cal O}_1 + C_2 {\cal O}_2 &=& a_1 {\cal O}_1 + 2 \tilde{\cal
O}_1 \no \\
\label {eqO3}
&=& a_2 {\cal O}_2 + 2 \tilde{\cal O}_2 \no
\eeqa
where

\beq
a_1 = C_1 + \xi C_2 \hs{15mm} a_2 = C_2 + \xi C_1 \hs{10mm} {\rm
with} \hs{5mm} \xi = \frac{1}{N_c} \ \label{eqO4}.
\eeq
Assuming factorization, the matrix element for ${\cal H}_{eff}$ for
$B^- \ra {D^*}^o \rho^-$ is written as

\beqa
<{D^*}^o\rho^-|C_1 {\cal O}_1 + C_2 {\cal O}_2|B^-> &=& \hs{2mm}
a_1 \hs{2mm} <\rho^-|\ol{d} \Gamma^{\rho} u |0> <{D^*}^o| \ol{c} \Gamma_{\rho}
b |B^-> \no \\
\cr
& & + \hs{1mm} a_2 \hs{2mm} <{D^*}^o| \ol{c} \Gamma^{\rho} u |0>
<\rho^-|\ol{d} \Gamma_{\rho} b |B^-> .
\label{eqO5}
\eeqa
 The contribution of the colour-octect currents vanishes in this
approximation
due to colour conservation.

In fact, we can rewrite ${\cal H}_{eff}$ in terms of "factorized
hadron operators" as
\cite{BSW}
\beq
{\cal H}_{eff} = \frac{G_F}{\sqrt{2}} \ V_{cb} V^*_{ud}
\{ a_1 [\ol{d} \Gamma^{\rho} u]_H [\ol{c} \Gamma_{\rho} b]_H
+ a_2 [\ol{c} \Gamma^{\rho} u]_H [\ol{d} \Gamma_{\rho} b]_H \}
\hs{2mm} + \hs{2mm} H.C. \label{eqO7}
\eeq
where the subscript $H$ stands for {\it hadronic} implying that the
Dirac bilinears of Eq.(12) be treated as interpolating fields for the
mesons and no futher Fierz-reordering need be done.

The contribution of the non factorizable term may have a significant
effect\footnote{For a nonfactorizable contribution to exclusive
nonleptonic weak decays, see for example Ref.\cite{HYCheng} }
: in fact,
this term might cancel a part of the factorizable one, more precisely
that proportional to $\xi C_1$ so that $a_2 \simeq C_2 (\xi \simeq
0)$.
Such cancellation seems to take place in charm two body decay
\cite{BSW}, however it is an  open question in $B$ meson two body
hadronic decays.
Moreover, the rule of discarding the contribution of the operators
with coloured currents while applying the vacuum saturation is
ambiguous and unjustified. It has, therefore, been suggested that
constants $a_1$ and $a_2$ be taken as free parameters.

\vs{3mm} $2^{\bullet})$ \hs{2mm}
While in the charm quark sector, a negative value of $a_2$
corresponding to $N_c \ra \infty $ is favoured by experimental data
\cite{BSW}, for the b quark sector, an analysis \cite{Neubert} of the
combined 'old data' from ARGUS and CLEO I yielded   two possible
solutions (positive/negative value) for $a_2$ when simultaneous fits
were made to Class I, II and III data :

\beqa
{\rm (i)} \hs{10mm} a_1 &=& 1.10 \pm 0.08, \hs{20mm} a_2 = 0.20 \pm
0.02  \no \\
{\rm (ii)} \hs{9mm} a_1 &=& 1.14 \pm 0.07, \hs{20mm} a_2 = -0.17 \pm
0.02  \no
\eeqa

However, the recent CLEO II data on the $B$ decays $\ol{B} \ra D
\pi(\rho), D^*\pi(\rho)$ exhibit a striking result
\cite{{Stone},{Browder},{CLEO94-5},{Yamamoto}}: the interference
between the two different amplitudes contributing to exclusive
two-body $B^-$ decays is evidently constructive, in contrast to what
is naively expected from the leading $1/N_c$ expansion \cite{BSW}.

Refs.\cite{Browder}, \cite{CLEO94-5} and \cite{Yamamoto} show
that the sign of $a_2/a_1$ is positive, contrary to what is found in
charm decays :
\beqa
& & \hs{2mm} 0.25 \pm 0.07\pm 0.05 \hs{30mm} {\rm \cite{Browder}} \no
\\
\frac{a_2}{a_1} \ \hs{5mm} &=&    \no \\
& & \hs{2mm} 0.23 \pm 0.04 \pm 0.04 \pm 0.10 \hs{15mm} {\rm
\cite{{CLEO94-5},{Yamamoto}}} \no
\eeqa

However Ref.\cite{GKKP} show that the question of the magitude and
the relative sign between $a_2$ and $a_1$  is still uncertain
when one standard deviation data are used.
When data are taken only with two standard deviation,
we found that $a_2$ is positive.

Faced with such uncertainties, we would like to suggest that
precision polarization measurements in $B \ra {D^o}^* \rho^-$ can
settle the question of  the sign and the magnitude of $a_2/a_1$.

\vs{6mm}
\hs{3mm} \Large{ } {\bf II. \hs{3mm} Generalities}
\vs{6mm} \normalsize

We start by describing
our  calculational procedure and parameters used.

\vs{2mm}
$1^{\bullet})$ We assume factorization \cite{BSW}.

\vs{2mm}
$2^{\bullet})$ We disregard the effect of final state interactions
(FSI) in this paper.
 On physical grounds FSI effects are expected to be small as the
produced quarks, being extremely relativistic, leave the strong
interaction region before hadronization so that FSI
between hadrons may not play a significant role.
Ref.\cite{KP} provides an estimate of the strong interaction phase
angles which turn out to be small.

\vs{2mm}
$3^{\bullet})$ Parameters we use are \cite{{Browder},{CLEO94-5}}

\hs{10mm} (a) \hs{2mm} $f_{\rho^+} = 212 \hs{2mm}MeV $ which is
measured from
$\tau \ra \nu_{\tau} \rho $ decay.

\hs{10mm} (b) \hs{2mm} $f_{D^*} = 220 \hs{2mm}MeV$ which comes from
theoretical estimates.

\vs{2mm}
$4^{\bullet})$ We introduce the following dimensionless quantities :

\beq
r_D \equiv \frac{m_{D^*}}{m_B} \
\hs{10mm}, \hs{10mm} t^2  \equiv \frac{q^2}{m_B^2} \
\eeq

\beq
k(t^2) = [(1 + r^2_D - t^2)^2 - 4 r^2_D]^{1/2}
\eeq
where $q^2 = (P_B - p_{D^*})^2$ is the square of $\rho$-meson momentum.

\vs{2mm}
$5^{\bullet})$ The other theoretical inputs in our calculation are
the hadronic form factors
for $B \ra D^*$ and $B \ra \rho$ transitions for which we use various
models.
Here we use the BSW definition of hadronic form factor \cite{BSW} :

\vs{2mm}
\hs{5mm}
$5^{\bullet}$-A) \hs{2mm}
$\ul{ B \ra D^* \rm \hs{2mm} transition \hs{2mm} in \hs{2mm}
spectator \hs{2mm} mode } $

\vs{2mm}
The spectator amplitude (${\cal A}$), multiplied by the parameter
$a_1$,
involves $B \ra D^*$ hadronic form factors which correspond to
 heavy  to heavy quark transition.
We shall consider the following two models of form factors : \vs{2mm}

\hs{10mm} (a) The set HQET I associated with exact heavy quark
symmetry
and used, for example, by Deandrea et al. \cite{Gatto} in their
analysis.
An extrapolation of the Isgur-Wise function $\xi(y)$ is made from the
symmetry point
using  an improved form of the relativistic oscillator model as
described in \cite{NR}.

\vs{2mm}
\hs{10mm} (b) The set HQET II associated with heavy quark symmetry
including mass corrections
as proposed by Neubert et al \cite{Neubert}.
We have made an interpolation of the entries in Table 4 of
\cite{Neubert}
to obtain the form factors needed.

\vs{2mm}
The decay amplitudes, apart from an overall factor, for the three
polarization states are given by
\beqa
{\cal A}_{LL}(q^2) &=& \frac{f_{\rho}}{m_B} \
\hs{1mm}(\frac{1 + r_D}{2 r_D} \ )
\hs{1mm}
\{(1 - r^2_D - t^2) \hs{2mm}A_1^{B D^*}(q^2) \hs{2mm} - \hs{2mm}
\frac{k^2(t^2)}{(1 + r_D)^2} \ \hs{2mm} A_2^{B D^*}(q^2) \}  \no \\
\cr
{\cal A}_{\pm\pm}(q^2) &=& \frac{f_{\rho}}{m_B} \ \hs{2mm} t \hs{2mm} (1 +
r_D) \{A_1^{B D^*}(q^2)
\hs{2mm} \mp \hs{2mm}
\frac{k(t^2)}{(1 + r_D)^2} \ V^{B D^*}(q^2) \}
\eeqa

\vs{2mm}
\hs{5mm}
$5^{\bullet}$-B) \hs{2mm}
$\ul{ B \ra \rho \rm \hs{2mm} transition \hs{2mm} in \hs{2mm}
colour-suppressed \hs{2mm} mode }$

\vs{2mm}
The colour-suppressed amplitudes (${\cal B}$), multiplied by the
parameter $a_2$,
involve $B \ra \rho$ hadronic form factors which correspond to
heavy  to light quark transition.

We shall consider here three sets of form factors :

\vs{2mm}
\hs{10mm} (a) The set BSW II which is a slightly modified version of
BSW I used in \cite{Neubert}.

The normalization at zero momentum transfer and the pole masses are unchanged,
however, the form factors $F_0^{B\rho}$ and $A_1^{B\rho}$
have a monopole type while the form factors $F_1^{B\rho},
A_0^{B\rho},
A_2^{B\rho} $ and $V^{B\rho}$ have a dipole type.

\vs{2mm}
\hs{10mm} (b) The set CDDFGN of Ref.\cite{CDD} used in the analysis
of Deandrea et al \cite{Gatto}, where the normalization of the heavy
to light transition form factors at zero momentum transfer has been estimated
in a
model combining chiral and heavy quark symmetry with mass
corrections.
The pole masses are as in Ref.\cite{BSW} and a monopole
type is used for all form factors.

\hs{10mm} (c) The set PB as suggested by Ref.\cite{Ball}, where
$A_1^{B\rho}$ is constant ( here we use $A_1^{B\rho} =
0.45$),  $V^{B\rho}$ has a monopole type with pole mass
$6.6 \hs{2mm}GeV$ and $V^{B\rho}(0) = 0.6$ and $A_2^{B\rho}$ is assumed to
be rising slowly  and parameterized as in Ref.\cite{Kamal}

\beq
\frac{A_2^{B\rho}(m_{D^*}^2)}{A_2^{B\rho}(0)} \ = 1 \hs{2mm} +  0.0222
\hs{3mm} \frac{m_{D^*}^2}{GeV^2}
\eeq
with $A_2^{B\rho}(0) = 0.4 $.

\vs{2mm}
The  colour-suppressed amplitudes for the three polarization states
are given by
\beqa
{\cal B}_{LL}(q^2) &=& \frac{f_{D^*}}{m_B} \
\hs{1mm} (\frac{1 + t}{2 \hs{2mm} t} \ ) \hs{1mm}
\{(1 - r_D^2 - t^2) \hs{2mm}A_1^{B \rho}(m_{D^*}^2) -
\frac{k^2(t^2)}{(1 + t)^2} \ \hs{2mm} A_2^{B \rho}(m_{D^*}^2) \}  \no  \\
\cr
{\cal B}_{\pm\pm}(q^2) &=& \frac{f_{D^*} }{m_B} \
\hs{1mm}  r_D \hs{2mm} (1 + t) \hs{1mm}
 \{A_1^{B \rho}(m_{D^*}^2) \mp \frac{k(t^2)}{(1 + t)^2} \
\hs{2mm} V^{B\rho}(m_{D^*}^2) \}.  \label{eq8}
\eeqa

\vs{2mm}
$6^{\bullet})$
The decay width in each of the three polarization states is

\beq
\Gamma_{\ld\ld}(B^- \ra {D^*}^o \rho^-) = \frac{G_F^2 m_B^5}{32 \pi}
\
|V_{cb}|^2 |V_{ud}|^2
\hs{1mm} k(t^2) \hs{1mm}
|a_1 {\cal A}_{\lambda \lambda}(q^2) + a_2 {\cal B}_{\lambda
\lambda}(q^2)|^2
\label{eq9}
\eeq

We now define the physical quantities which we will discuss in this
paper :

\vs{2mm}
\hs{10mm}
(a) Longitudinal polarization.

\beqa
\rho_L \hs{2mm} &\equiv& \hs{2mm} \frac{ \Gamma_{LL}}{ \Gamma} \
 =  \frac{ \Gamma_{LL}}{ \sum_{\ld} \hs{2mm} \Gamma_{\ld\ld} } \no \\
\cr
& = & \hs{2mm} \frac{| \A_{LL} \hs{2mm} + \hs{2mm} \zeta \hs{2mm}
\B_{LL} |^2}
{\sum_{\ld} \hs{2mm} | \A_{\ld\ld} \hs{2mm} + \hs{2mm} \zeta \hs{2mm}
\B_{\ld\ld} |^2} \
\eeqa

\hs{10mm}
(b) Left-right asymmetry in transverse polarization.

\beqa
\A_{LR} \hs{2mm} &\equiv & \hs{2mm} \frac{ \Gamma_{--} \hs{2mm} -
\hs{2mm} \Gamma_{++}}
{ \Gamma_{--} \hs{2mm} + \hs{2mm} \Gamma_{++}} \no \\
& = & \hs{2mm} \frac{ | \A_{--} \hs{2mm} + \hs{2mm} \zeta \hs{2mm}
\B_{--} |^2
\hs{2mm} - \hs{2mm} | \A_{++} \hs{2mm} + \hs{2mm} \zeta \hs{2mm}
\B_{++} |^2 }
{ | \A_{--} \hs{2mm} + \hs{2mm} \zeta \hs{2mm} \B_{--} |^2
\hs{2mm} + \hs{2mm} | \A_{++} \hs{2mm} + \hs{2mm} \zeta \hs{2mm}
\B_{++} |^2 }
\eeqa
with $\zeta \equiv a_2/a_1$.

\vs{6mm}
\hs{3mm} \Large{ } {\bf III . \hs{3mm} Analysis}
\vs{6mm} \normalsize

We now analyze how  sensitive are the longitudinal fraction
and the left-right transverse asymmetry to the magnitude and to the
sign of $a_2/a_1$
, in the zero-width approximation and a finite-width calculation of
the $\rho$-meson respectively.

\vs{3mm}
\hs{3mm} \large{ } {\bf II-a. \hs{3mm} Zero-width approximation  in
${\bf \rho}$-meson(ZW)}. \normalsize
\vs{3mm}

Within the zero-width approximation for the $\rho$-meson, quantities
defined in Eq.(14) become :

\beq
q^2 \equiv m_{\rho}^2 \hs{10mm}, \hs{10mm} t = \frac{m_{\rho}}{m_B} \
\eeq
and we introduce the superscript "0" to signify quantites in the
zero-width approximation.

The decay width in the three polarization states is given by :

\beq
\Gamma_{\ld\ld}^0(B^- \ra {D^*}^o \rho^-) = \frac{G_F^2 m_B^5}{32
\pi} \
|V_{cb}|^2 |V_{ud}|^2
\hs{1mm} k(m_{\rho^2}) \hs{1mm}
|a_1 {\cal A}_{\lambda \lambda}(m_{\rho}^2) + a_2 {\cal B}_{\lambda
\lambda}(m_{\rho}^2)|^2
\eeq

\vs{2mm}

\hs{10mm}
$1^{\bullet})$. \hs{2mm}
The longitudinal fraction in the zero-width approximation is :

\beq
\rho_L^0  =  \frac{| \A_{LL}(m_{\rho}^2) \hs{2mm} +
\hs{2mm} \zeta \hs{2mm} \B_{LL}(m_{\rho}^2) |^2}
{\sum_{\ld} \hs{2mm} | \A_{\ld\ld}(m_{\rho}^2) \hs{2mm} +
\hs{2mm} \zeta \hs{2mm} \B_{\ld\ld}(m_{\rho}^2) |^2} \
\eeq

In Table 1 corresponding to each models used,
we show results of the relative amount of the longitudinal
polarization component $\rho_L^0$ corresponding to four values of
$\zeta$ : $\zeta = \pm 0.25$ and $\zeta = \pm 0.20$.

Figure 2-a and  2-b  illustrate the sensitivity of this quantity
to the relative sign and the magnitude of $\zeta = a_2/a_1$.

\vs{3mm}
\begin{table}[thb]
\begin{center}
\begin{tabular}{|c||c|c|c|c||c|c|c|c||}
\hline
$\rho_L^0$  & \multicolumn{4}{c||}{ HQET I } &
  \multicolumn{4}{c||}{ HQET II } \\
\hline
\hline
$ \zeta = a_2/a_1 $ & -0.25 & -0.20 & 0.20 & 0.25  &
 -0.25 & -0.20 & 0.20 & 0.25  \\
\hline
\hline
 BSW II  &  0.924 & 0.916 & 0.854 & 0.848  &  0.925 & 0.916 & 0.848 &
0.841  \\
\hline
 CDDFGN  &  0.913 & 0.909 & 0.854 & 0.846  &  0.908 & 0.904 & 0.851 &
0.844  \\
\hline
 PB   &  0.928 & 0.919 & 0.856 & 0.850  &  0.933 & 0.921 & 0.850
& 0.844   \\
\hline
\end{tabular}

\vs{3mm}
Table 1. \\
The longitudinal polarization fraction in the $\rho$-meson zero-width
approximation
\end{center}
\end{table}

\vs{2mm}
\hs{10mm}
$2^{\bullet})$. \hs{2mm}
Next, we consider the transverse left-right helicity asymmetry
in the zero-width approximation :

\beq
\A_{LR}^0 =
\frac{ | \A_{--}(m_{\rho}^2) \hs{2mm} + \hs{2mm} \zeta \hs{2mm}
\B_{--}(m_{\rho}^2) |^2
\hs{2mm} - \hs{2mm} | \A_{++}(m_{\rho}^2) \hs{2mm} + \hs{2mm} \zeta
\hs{2mm} \B_{++}(m_{\rho}^2) |^2 }
{ | \A_{--}(m_{\rho}^2) \hs{2mm} + \hs{2mm} \zeta \hs{2mm}
\B_{--}(m_{\rho}^2) |^2
\hs{2mm} + \hs{2mm} | \A_{++}(m_{\rho}^2) \hs{2mm} + \hs{2mm} \zeta
\hs{2mm} \B_{++}(m_{\rho}^2) |^2 }
\eeq

The amount of left-right asymmetry in transverse polarization, in the
zero-width approximation, corresponding to values of $\zeta$, \hs{1mm} $\zeta =
\pm 0.25$ and $\zeta = \pm 0.20$,  is shown in Table 2 and the
sensitivity  to the relative sign and the magnitude of $\zeta =
a_2/a_1$ for each models is shown in Figures 3-a and 3-b.

\vs{3mm}
\begin{table}[thb]
\begin{center}
\begin{tabular}{|c||c|c|c|c||c|c|c|c||}
\hline
$A_{LR}^0 $
& \multicolumn{4}{c||} { HQET I }
& \multicolumn{4}{c||} {HQET II}  \\
\hline
\hline
$ \zeta = a_2/a_1 $ &  -0.25 & -0.20 & 0.20 & 0.25  & -0.25 & -0.20 &
0.20 & 0.25  \\
\hline
\hline
BSW II  &   0.489 & 0.562 & 0.819 & 0.833  &  0.753 & 0.789 & 0.910 &
0.916  \\
\hline
 CDDFGN  &  0.077 & 0.265 & 0.915 & 0.938 &  0.361 & 0.527 & 0.973 &
0.984  \\
\hline
 PB  & 0.216  & 0.396 & 0.850 & 0.867  &  0.583 & 0.693 & 0.927 &
0.934 \\
\hline
\end{tabular}

\vs{3mm}
Table 2. \\
The left-right asymmetry of the transverse polarizations \\
in the $\rho$-meson zero-width approximation
\end{center}
\end{table}

Table 1 and  Figures 2-a and 2-b  show that the longitudinal fraction
does not depend drastically upon the form factor models in the region
we consider ($ |\zeta| \leq 0.3 $ ).
However, the relative difference  between two  points $\zeta = \pm
0.2$ or $\pm 0.25$ reached about $6 \sim 8$ percent, which might be
distinguished  in future experiments.

We remark that if the longitudinal fraction is greater than 0.88,
the parameter $a_2$ is negative
and if less than 0.88, $a_2$ is  positive.

Figures 3-a and 3-b show  that the left-right helicity asymmetry
fraction varied
considerably model to model.

Because the value of the left-right asymmetry between two relative
points varies from 0.12 to 0.86,
the measurement of ${\cal A}_{LR}$ would be more sensitive to the
sign of $\zeta$.
In the zero-width approximation of the $\rho$-meson, we conclude :
 (i) \hs{2mm}with the HQET I model of $B \ra D^*$,
if ${\cal A}_{LR}^0$ is greater than 0.74, $a_2$ is positive
and if ${\cal A}_{LR}^0$ is smaller than 0.74, $a_2$ is negative.
\hs{3mm} (ii) \hs{2mm} with the HQET II model,
if ${\cal A}_{LR}^0$ is greater than 0.87, $a_2$ is positive
and if ${\cal A}_{LR}^0$ is smaller than 0.87, $a_2$ is negative.

\vs{6mm}
\hs{3mm} \large{ } {\bf II-b. \hs{3mm} {\bf $\rho$}-meson Non
zero-width calculation
{\bf (FW)}. \normalsize
\vs{4mm}

In the above calculation, $\rho$-meson was assumed to have zero width
which is certainly a poor approximation.
The final state $\rho$-meson has a width of about $150 \hs{2mm}MeV$,
and this rather wide resonance increases the effective final-state
phase space.
If we take the finite $\rho$-width into account,
we have to smear the rate given in Eq.(22) over $t^2$ with a
Breit-Wigner measure \cite{Pham} :
\beq
\delta(t^2 - r_{\rho}^2)  \ra \frac{1}{\pi} \ \hs{2mm} \frac{r_{\rho}
\gamma_{\rho} }{(t^2 - r_{\rho}^2)^2 + r_{\rho}^2 \gamma_{\rho}^2 } \
\equiv BW(t^2)
\eeq
where we introduce the dimemsionless variables : $r_{\rho} = m_{\rho}/m_B$
and
$\gamma_{\rho} = \Gamma_{\rho}/m_B$,
($\Gamma_{\rho} = 151.5 \hs{2mm} MeV$ is the total $\rho$-meson width).

\vs{2mm}
The decay width in the three polarization states in Eq.(22) gets
modified to
\beqa
& & k(m_{\rho}^2) \hs{2mm} | {\cal A}_{\ld\ld}(m_{\rho}^2) \hs{2mm} +
\hs{2mm} \zeta \hs{2mm} {\cal B}_{\ld\ld}(m_{\rho}^2) |^2 \no \\
\cr
& &
\Lra \hs{5mm}
\int_{\frac{4 m_{\pi}^2}{m^2_B} \ }^{(1 - r_D)^2}
dt^2
\hs{2mm} BW(t^2) \hs{2mm} k(t^2) \hs{2mm} | {\cal A}_{\ld\ld}(q^2)
\hs{2mm} + \hs{2mm} \zeta \hs{2mm} {\cal B}_{\ld\ld}(q^2)|^2
\eeqa

\vs{2mm} $1^{\bullet})$. \hs{2mm}
In the Table 3, we tabulate the longitudinal polarization fraction,
$\rho_L$, calculated with a finite width $\rho$-meson in  each model.
And in Figure 4-a and 4-b,
we show the sensitivity of the longitudinal polarization fraction to
$\zeta$.

\vs{3mm}
\begin{table}[thb]
\begin{center}
\begin{tabular}{|c||c|c|c|c||c|c|c|c||}
\hline
$\rho_L$ (FW)  & \multicolumn{4}{c||}{ HQET I } &
  \multicolumn{4}{c||}{ HQET II } \\
\hline
\hline
$ \zeta = a_2/a_1 $ & -0.25 & -0.20 & 0.20 & 0.25  &
 -0.25 & -0.20 & 0.20 & 0.25  \\
\hline
\hline
 BSW II  &  0.902 & 0.895 & 0.839 & 0.833  &  0.903 & 0.894 & 0.834 &
0.827  \\
\hline
 CDDFGN  &  0.890 & 0.887 & 0.839 & 0.833  &  0.896 & 0.890 & 0.831 &
0.824  \\
\hline
 PB   &  0.899 & 0.893 & 0.843 & 0.838  &  0.903 & 0.894 & 0.837
& 0.832   \\
\hline
\end{tabular}

\vs{5mm}
Table 3. \\
The longitudinal polarization fraction in the $\rho$-meson
finite-width calculation
\end{center}
\end{table}

\vs{2mm} $2^{\bullet})$. \hs{2mm}
Table 4 shows the relative amount of the left-right asymmetry of the
transverse
polarization in a finite-width  $\rho$-meson calculation.
And in Figures 5-a and 5-b, we show the sensitivity to the sign
and the magnitude of $\zeta$ for each model.

\vs{3mm}
\begin{table}[thb]
\begin{center}
\begin{tabular}{|c||c|c|c|c||c|c|c|c||}
\hline
$A_{LR}$ (FW)
& \multicolumn{4}{c||} { HQET I }
& \multicolumn{4}{c||} {HQET II}  \\
\hline
\hline
$ \zeta = a_2/a_1 $ &  -0.25 & -0.20 & 0.20 & 0.25  & -0.25 & -0.20 &
0.20 & 0.25  \\
\hline
\hline
BSW II  &   0.528 & 0.580 & 0.796 & 0.810  &  0.763 & 0.789 & 0.893 &
0.899  \\
\hline
 CDDFGN  &  0.200 & 0.341 & 0.886 & 0.909 &  0.460 & 0.580 & 0.954 &
0.965  \\
\hline
 PB  & 0.350  & 0.462 & 0.827 & 0.843  &  0.656 & 0.721 & 0.909 &
0.917 \\
\hline
\end{tabular}

\vs{5mm}
Table 4. \\
The transverse left-right asymmetry in the finite-width $\rho$-meson
calculation
\end{center}
\end{table}

As seen from Figures 4 and 5, both the longitudinal polarization
fraction and the transverse left-right asymmetry vary more smoothly
in the finite-width $\rho$-meson calculation
than in the zero-width one, due to the smearing  effect.
For a finite width $\rho$-meson, $a_2$ has a positive value  if the
longitudinal polarization fraction is less than 0.86 and negative if
the longitudinal polarization fraction is greater than 0.86.

As for the transverse left-right asymmetry, our results show that if
$\A_{LR}$ is greater than 0.86, $a_2$ is positive and if $\A_{LR}$
is smaller than 0.72, $a_2$ is negative,
depending on heavy to heavy transition form factor models considered.

\newpage
\vs{6mm}
\hs{3mm} \Large{ } {\bf III \hs{3mm} Discussion}
\vs{4mm} \normalsize

\vs{2mm} $1^{\bullet})$. \hs{2mm}
As shown in Figures 4 and 5, we find that the smearing effect of a
finite-width $\rho$-meson is considerable :
The longitudinal component of the decay width is changed about $15
\%$ and the transverse components of decay width are changed by more
than $20 \%$.
The longitudinal fraction is shifted by about $2 \%$ and varies more
smoothly than for a zero-width $\rho$-meson.
The left-right asymmetry of the transverse polarization is much more
affected by the smearing effect as shown in Figures 5-a and 5-b.

\vs{2mm} $2^{\bullet})$. \hs{2mm}
The longitudinal fraction $\rho_L$ for reasonable value of $\zeta$
is only weakly model dependent.
In the non-zero $\rho$-meson width calculation
we obtain the following results :

\vs{2mm}
\hs{10mm} 1) \hs{2mm}
If $\rho_L$ is large, $\rho_L > 0.86$,
negative values of $a_2/a_1$ are favoured  and
for $\rho_L < 0.86$, positive values emerge.
We must keep in mind that the value $\rho_L = 0.86$ at $a_2 = 0$ depends
only on the spectator amplitude or equivalently of
the heavy to heavy transition and it is more reliable.

\vs{2mm}
\hs{10mm} 2) \hs{2mm}
The model dependence of $\rho_L$ is moderate :
for instance with  $a_2/a_1 = 0.2$ we obtain
$\rho_L = 0.831 - 0.843$ from Table 3, for $a_2/a_1 = - 0.20$,
we get $\rho_L = 0.887 - 0.895$.

\vs{2mm} $3^{\bullet})$. \hs{2mm}
The transverse left-right asymmetry $\A_{LR}$ appears to be
more model dependent than $\rho_L$ especially for negative $a_2/a_1$.
In the non-zero $\rho$-meson width calculation the results are the following :

\vs{2mm}
\hs{10mm} 1) \hs{2mm}
If $\A_{LR}$ is larger than $0.72$ with HQET I and 0,86 with HQET II,
positive values of $a_2/a_1$ are favoured.
Figures 5-a and 5-b also depend only on heavy to heavy transition and
are reliable even if the model dependence is more important than for $\rho_L$.

\vs{2mm}
\hs{10mm} 2) \hs{2mm}
For $a_2/a_1$ positive the model dependence remains moderate.
For instance with $a_2/a_1 = 0.2$ we obtain, from Table 4,
$\A_{LR} = 0.796 - 0.954$.
However
for $a_2/a_1$ negative, the model dependence is very important
and at $a_2/a_1 = - 0.2$ we obtain
$\A_{LR} = 0.341 - 0.789$.
With very large negative value of $a_2/a_1$,
$\A_{LR}$ may even become negative.

\vs{2mm} $4^{\bullet})$. \hs{2mm}We would like to emphasize
the difference between our approach and
the ones adopted by others
\cite{{Neubert},{Browder},{CLEO94-5},{Yamamoto}}.
These authors determine the size and the sign of $a_2/a_1$ from a
global fit
to the ratios of Class III to Class I rates as defined in
\cite{{Neubert},{Browder},{CLEO94-5},{Yamamoto}}.
In our approach, we propose to determine the magnitude and the sign
of $a_2/a_1$ by using the sensitivity of the longitudinal
polarization fraction and  the left-right asymmetry in transverse
polarization fraction of Class III process alone.

Therefore measurements of the longitudinal
polarization fraction and the transverse left-right helicity
asymmetry of $B^- \ra \rho^- {D^*}^o$ decay mode seem to be important.

%
%
%
%
\newpage
\vspace{1cm}
\hspace{1cm} \Large{} {\bf Acknowledgements}    \vspace{0.6cm}

\normalsize{
I am grateful to M. Gourdin, A. N. Kamal and specially X. Y. Pham
for collaboration and discussions.
Without their contributions and encouragement this report would not have been
written.
I also acknowledge helpful suggestions and correspondence with K. Honschied and
H. Yamamoto.
I would like to thank the Laboratoire de Physique Th\'eorique et Hautes
Energies in Paris for their hospitality and P. Gondolo for helping computer
utilization.
Finally I  thank the Commissariat \`a l'Energie Atomique of France for the
award of a fellowship
and especially G. Cohen-Tannoudji and J. Ha\"{\i}ssinski for  their
encouragements.

}

\newpage
%

\newpage
\section*{Figure captions}
\normalsize
\vspace{0.5cm}

\begin{enumerate}

\item
{\bf Fig. 1} : \hs{2mm}
Feynman diagrams for $B^- \ra {D^*}^o \rho^-$ decay. \\
(a) external spectator diagram. \\
(b) internal spectator (colour-suppressed) diagram.

\item
\begin{enumerate}
\item
{\bf Fig. 2-a} : \hs{2mm}
The longitudinal polarization fraction $\Gamma_L/\Gamma$

in  $B^- \ra {D^*}^o
\rho^-$ decay  within zero-width approximation for the $\rho$ meson,
\\
(1) the solid line corresponds to BSW II model in $B \ra \rho$
transition, \\
(2) the dotted line corresponds to CDDFGN model in $B \ra \rho$
transition, \\
(3) the dash-dotted line corresponds to PB model in
$B \ra \rho$ transition \\
with HQET I model in $B \ra D^*$ transition. \\
\item
{\bf Fig. 2-b} : \hs{2mm}
Same as Fig. 2-a with HQET II model in $B \ra D^*$ transition.

\end{enumerate}

\item
\begin{enumerate}
\item
{\bf Fig. 3-a} : \hs{2mm}
The left-right helicity asymmetry of the transverse polarization
fraction ${\cal A}_{LR}$
in $B^- \ra {D^*}^o\rho^-$ decay
within zero width approximation for the $\rho$ meson. \\
(1) the solid line corresponds to BSW II model in $B \ra \rho$
transition. \\
(2) the dotted line corresponds to CDDFGN model in $B \ra \rho$
transition. \\
(3) the dash-dotted line corresponds to PB model in
$B \ra \rho$ transition \\
with HQET I model in $B \ra D^*$ transition.

\item
{\bf Fig. 3-b} : \hs{2mm}
Same as Fig. 3-a with HQET II model in $B \ra D^*$ transition.

\end{enumerate}

\item
{\bf Fig. 4-a and 4-b} : \hs{2mm}
Same as Figures 2-a and 2-b with the finite-width $\rho$ meson.

\item
{\bf Fig. 5-a and 5-b} : \hs{2mm}
Same as Figures 3-a and 3-b with the finite-width  $\rho$ meson.
\end{enumerate}

\newpage
\section*{Table captions}
\normalsize
\vspace{0.5cm}

\begin{enumerate}

\item
{\bf Table 1} : \\
The longitudinal polarization fraction $\Gamma_L/\Gamma$
in  $B^- \ra {D^*}^o
\rho^-$ decay  at $a_2/a_1$ = -0.25, -0.20, 0.20 and 0.25
with HQET I, HQET II models for $B \ra D^*$ transition
and BSW II, CDDFGN, PB models for $B \ra \rho$ transition
with the zero-width $\rho$ meson.

\item
{\bf Table 2} : \\
The left-right asymmetry of the polarization fraction
${\cal A}_{LR}$ in  $B^- \ra {D^*}^o \rho^-$ decay
at $a_2/a_1$ = -0.25, -0.20, 0.20 and 0.25
with  HQET I, HQET II models for $B \ra D^*$ transition
and BSW II, CDDFGN, PB models for $B \ra \rho$ transition
with the zero-width $\rho$ meson

\item
{\bf Table 3} : \\
The longitudinal polarization fraction $\Gamma_L/\Gamma$
in  $B^- \ra {D^*}^o
\rho^-$ decay  at $a_2/a_1$ = -0.25, -0.20, 0.20 and 0.25
with  HQET I, HQET II models for $B \ra D^*$ transition
and BSW II, CDDFGN, PB models for $B \ra \rho$ transition
with the finite-width $\rho$meson.

\item
{\bf Table 4} : \\
The left-right asymmetry of the polarization fraction
${\cal A}_{LR}$ in $B^- \ra {D^*}^o \rho^-$ decay
at $a_2/a_1$ = -0.25, -0.20, 0.20 and 0.25
with  HQET I, HQET II models for $B \ra D^*$ transition
and BSW II, CDDFGN, PB models for $B \ra \rho$ transition
with the finite-width  $\rho$ meson.

\end{enumerate}



\begin{thebibliography}{99}
%
%
\bibitem{Neubert}
M.~Neubert, V.~Rieckert, B.~Stech and Q.P.~Xu, in  $\ul{Heavy
\hs{2mm} Flavours}$,
 Eds. A. J. Buras and M. Lindner, ( World Scientific, Singapore, 1992
).
%
\bibitem{Stone}
S. ~Stone, Talk presented at the 5th International symposium on Heavy
Flavor Physics
, Montr\'eal, 1993 ( to  be published ).
%
\bibitem{KP} A. N. ~Kamal and T. N.~Pham, {\em Phys. Rev.} {\bf D 50}
395 (1994).
%
\bibitem{Browder}
T. E.~Browder, K.~Honscheid  and S.~Playfer, CLEO Report No. CLNS
93/1261,UH-551-778-93, OHSTPY-HEP-E 93-108, HEPSY 93-10
(to be appeared in $\ul{B \hs{2mm} decays}$, 2nd edition,
Ed. by S. Stone, World Scientific, Singapore).
%
\bibitem{GKKP}
M. ~Gourdin, A. N. ~Kamal, Y. Y. ~Keum and X. Y. ~Pham,
{\em Phys. Lett.} {\bf B 333}, 507 (1994).
%
\bibitem{CLEO94-5}
M. S. Alam et al., CLEO Collaboration, {\em Phys. Rev.} {\bf D 50},
43 (1994).
%
\bibitem{Yamamoto}
H. ~Yamamoto, Harvard Report No HUTP-93/A039.
%
\bibitem{BSW}
M. ~Wirbel, B. ~Stech and M. ~Bauer, {\em Z. Phys.} {\bf C29} 637
(1985); \\
M. ~Bauer, B. ~Stech and M. ~Wirbel, {\em Z. Phys.} {\bf C34} 103
(1987);
ibid, {\em Z. Phys.} {\bf C42} 671 (1989)
%
\bibitem{Gatto} A. ~Deandrea, N. ~Di Bartolomeo, R. ~Gatto and G.
{}~Nardulli, {\em Phys. Lett.} {\bf B 318}, 549 (1993).
%
\bibitem{NLL1}
G. Altarelli, G. Curci, G. Martinelli and S. Petrarca, {\em Nucl.
Phys.} {\bf B 187}, 461 (1981); {\em Phys. Lett.} {\bf B 99},
(1981).
%
\bibitem{NLL2}
A. Buras, P. H. Weisz, {\em Nucl. Phys.} {\bf B 333}, 66 (1990). \\
A. Buras, M. Jamin, M. E. Lautenbacher and P. H. Weisz, {\em Nucl.
Phys.} {\bf B 370}, 69 (1990).
%
\bibitem{LLA}
M. K. Gaillard and B. W. Lee, {\em Phys. Rev. Lett.} {\bf 33}, 108
(1974); \\
G. Altarelli and L. Maiani, {\em Phys. Lett.} {\bf B 32}, 351 (1974).
%
\bibitem{Ruckl}
R. R\"uckl, Preprint MPT-PH/36/89.
%
%
%
\bibitem{Fierz}
L. B. Okun, ' Leptons and Quarks' ( North Holland, 1982); \\
J. F. Donoghue, E. Golowich and B. R. Holstein, 'Dynamics of the
Standard Model' P 217 (Cambridge Uni
versity Press, 1992); \\
A. ~Deandrea, N. ~Di Bartolomeo, R. ~Gatto and G. ~Nardulli, see
Ref.\cite{Gatto}; \\
A. Khodjamirian and R. R\"uckl, MPI Report No. MPI-PhT/94-26, LMU
05/94 , Feurary 1994
%
\bibitem{HYCheng}
Hai-Yang ~Cheng, {\em Phys. Lett.} {\bf B 335}, 428 (1994) ; \\
J. M. ~Soares, Preprint TRI-PP-94-78, September, 1994,
hep-ph/9409443.
%
%
\bibitem{NR}
M.~Neubert and V.~Rieckert, {\em Nucl. Phys.} {\bf B285}, 97 (1992).
%
\bibitem{CDD}
R. ~Casalbuoni, A. ~Deandrea, N. Di ~Bartolomeo, F. ~Feruglio, R. ~Gatto
and G. ~Nardulli, {\em Phys. Lett.} {\bf B 292}, 371(1992); ibid {\bf B
299}, 139(1993).
%
\bibitem{Ball}
P. ~Ball, {\em Phys. Rev.} {\bf D 48}, 3190 (1993).
%
\bibitem{Kamal}
A. N. ~Kamal and A. N. Santra, Preprint Alberta Thy-27-94;
hep-ph/9409364,
September, 1994.
%
\bibitem{Pham}
X. Y. Pham and X. C. Vu, {\em Phys. Rev.} {\bf D46}, 261 (1992); \\
T. N. Pham, {\em Phys. Rev.} {\bf D46}, 2976 (1992); \\
M. Gourdin, A. N. Kamal, Y. Y. Keum and X. Y. Pham,
LPTHE Report No. PAR/LPTHE/94-30, 1994; Preprint hep-ph/9407400.



\end{thebibliography}
\end{document}